\documentclass[lettersize,journal]{IEEEtran}
\usepackage{amsmath,amsfonts}
\usepackage{algorithmic}
\usepackage{algorithm}
\usepackage{array}
\usepackage[caption=false,font=normalsize,labelfont=sf,textfont=sf]{subfig}
\usepackage{textcomp}
\usepackage{stfloats}
\usepackage{url}
\usepackage{verbatim}
\usepackage{graphicx}
\usepackage{cite}
\usepackage{mathtools,amssymb,tikz,bm,siunitx, booktabs, subcaption}
\usetikzlibrary{positioning, shapes.geometric}            %
\usepackage{pgfplots}
\usetikzlibrary{arrows,calc,fit,matrix,positioning,shapes,shadows,trees,mindmap,tikzmark,arrows,arrows.meta,angles,quotes,babel,spy,decorations.markings}
\usetikzlibrary{external}
\usepackage{caption}
\usepackage{graphicx}
\usepackage{subcaption}

\newcommand{\Cs}{$\mathcal{C}_1$}       %
\newcommand{\Cb}{$\mathcal{C}_2$}       %
\newcommand{\Systemname}{Elena-SNN}
\newcommand{\Discriminator}{$^\star$}

\definecolor{kit-green100}{rgb}{0,.59,.51}
\definecolor{kit-green70}{rgb}{.3,.71,.65}
\definecolor{kit-green50}{rgb}{.50,.79,.75}
\definecolor{kit-green30}{rgb}{.69,.87,.85}
\definecolor{kit-green15}{rgb}{.85,.93,.93}
\definecolor{KITgreen}{rgb}{0,.59,.51}

\definecolor{KITpalegreen}{RGB}{130,190,60}
\colorlet{kit-maigreen100}{KITpalegreen}
\colorlet{kit-maigreen70}{KITpalegreen!70}
\colorlet{kit-maigreen50}{KITpalegreen!50}
\colorlet{kit-maigreen30}{KITpalegreen!30}
\colorlet{kit-maigreen15}{KITpalegreen!15}

\definecolor{KITblue}{rgb}{.27,.39,.66}
\definecolor{kit-blue100}{rgb}{.27,.39,.67}
\definecolor{kit-blue70}{rgb}{.49,.57,.76}
\definecolor{kit-blue50}{rgb}{.64,.69,.83}
\definecolor{kit-blue30}{rgb}{.78,.82,.9}
\definecolor{kit-blue15}{rgb}{.89,.91,.95}

\definecolor{KITyellow}{rgb}{.98,.89,0}
\definecolor{kit-yellow100}{cmyk}{0,.05,1,0}
\definecolor{kit-yellow70}{cmyk}{0,.035,.7,0}
\definecolor{kit-yellow50}{cmyk}{0,.025,.5,0}
\definecolor{kit-yellow30}{cmyk}{0,.015,.3,0}
\definecolor{kit-yellow15}{cmyk}{0,.0075,.15,0}

\definecolor{KITorange}{rgb}{.87,.60,.10}
\definecolor{kit-orange100}{cmyk}{0,.45,1,0}
\definecolor{kit-orange70}{cmyk}{0,.315,.7,0}
\definecolor{kit-orange50}{cmyk}{0,.225,.5,0}
\definecolor{kit-orange30}{cmyk}{0,.135,.3,0}
\definecolor{kit-orange15}{cmyk}{0,.0675,.15,0}

\definecolor{KITred}{rgb}{.63,.13,.13}
\definecolor{kit-red100}{cmyk}{.25,1,1,0}
\definecolor{kit-red70}{cmyk}{.175,.7,.7,0}
\definecolor{kit-red50}{cmyk}{.125,.5,.5,0}
\definecolor{kit-red30}{cmyk}{.075,.3,.3,0}
\definecolor{kit-red15}{cmyk}{.0375,.15,.15,0}

\definecolor{KITpurple}{RGB}{160,0,120}
\colorlet{kit-purple100}{KITpurple}
\colorlet{kit-purple70}{KITpurple!70}
\colorlet{kit-purple50}{KITpurple!50}
\colorlet{kit-purple30}{KITpurple!30}
\colorlet{kit-purple15}{KITpurple!15}

\definecolor{KITcyanblue}{RGB}{80,170,230}
\colorlet{kit-cyanblue100}{KITcyanblue}
\colorlet{kit-cyanblue70}{KITcyanblue!70}
\colorlet{kit-cyanblue50}{KITcyanblue!50}
\colorlet{kit-cyanblue30}{KITcyanblue!30}
\colorlet{kit-cyanblue15}{KITcyanblue!15}

\usepackage{wrapfig}
\usepackage[font={footnotesize}]{caption}           %
\DeclareMathAlphabet{\mathcal}{OMS}{cmsy}{m}{n}     %

\usepackage[colorlinks=true,bookmarks=false,linkcolor=.,citecolor=.,urlcolor=blue]{hyperref}
\definecolor{cR1}{rgb}{0,0,0}%
\definecolor{cR2}{RGB}{0,0,0}%
\definecolor{cR12}{RGB}{0,0,0}%

\newcommand{\RevI}[1]{\textcolor{cR1}{#1}}
\newcommand{\RevII}[1]{\textcolor{cR2}{#1}}
\newcommand{\RevIII}[1]{\textcolor{cR12}{#1}}

\begin{document}

\title{Spiking Neural Belief Propagation Decoder for Short Block Length LDPC Codes}

\author{Alexander~von~Bank,~\IEEEmembership{Graduate~Student~Member,~IEEE}, Eike-Manuel~Edelmann,~\IEEEmembership{Graduate~Student~Member,~IEEE}, Sisi~Miao,~\IEEEmembership{Student~Member,~IEEE}, Jonathan~Mandelbaum,~\IEEEmembership{Graduate~Student~Member,~IEEE}, Laurent~Schmalen,~\IEEEmembership{Fellow,~IEEE,}
\vspace{-0.6cm}
\thanks{The authors are with the Communications Engineering Lab,
Karlsruhe Institute of Technology, 76187 Karlsruhe, Germany. (e-mail:\{alexander.bank, eike-manuel.edelmann, sisi.miao, jonathan.mandelbaum, laurent.schmalen\}@kit.edu). \\
This work has received funding from the European Research Council (ERC) under the European Union's Horizon 2020 research and innovation program (grant agreement No. 101001899). Parts of this work were carried out in the framework of the CELTIC-NEXT project AI-NET-ANTILLAS (C2019/3-3) (grant agreement 16KIS1316) and within the project Open6GHub (grant agreement 16KISK010) funded by the German Federal Ministry of Education and Research (BMBF). }%
}

\markboth{IEEE COMMUNICATIONS LETTERS,~Vol.~XX, No.~?, YY~2024}%
{Shell \MakeLowercase{\textit{et al.}}: A Sample Article Using IEEEtran.cls for IEEE Journals}

\maketitle

\begin{abstract}
Spiking neural networks (SNNs) are neural networks that enable energy-efficient signal processing due to their event-based nature. 
This paper proposes a novel decoding algorithm for low-density parity-check (LDPC) codes that integrates SNNs into belief propagation (BP) decoding by approximating the check node update equations using SNNs.
For the (273,191) and (1023,781) finite-geometry LDPC code, the proposed decoder outperforms sum-product decoder at high signal-to-noise ratios (SNRs).
The decoder achieves a similar bit error rate to normalized sum-product decoding with successive relaxation.
Furthermore, the novel decoding operates without requiring knowledge of the SNR, making it robust to SNR mismatch.
\end{abstract}

\begin{IEEEkeywords}
Spiking Neural Networks, Channel Codes, Decoder, Message Passing, Belief Propagation Decoding, LDPC Codes, Neuromorphic, Event-based Computing.
\end{IEEEkeywords}

\vspace*{-0.3cm}
\section{Introduction}
\IEEEPARstart{N}{euromorphic} computing has gained much attention in recent years as CPUs are approaching their physical limits of operation speed and efficiency \cite{FSTNP2017}.
Spiking neural networks (SNNs) mimic the behavior of the human brain, which is a highly efficient computational device that consumes only $\SI{20}{\watt}$ of power to solve several complex tasks concurrently \cite{RJP2019}.
Hence, implementing SNNs on neuromorphic hardware is particularly interesting since it promises energy-efficient signal processing due to its event-based computing \cite{YK2020}. 
Furthermore, SNNs can be implemented on various types of neuromorphic hardware, enabling computation in the analog, digital, or photonic domain \cite{FSTNP2017,Intel2021,PBXKSSWLMS22}.
\RevI{An SNN-based equalizer was emulated on an FPGA and compared to benchmark equalizers in~\cite{MNHW24}, demonstrating the superior energy efficiency of SNNs.}

Low-density parity-check (LDPC) codes are forward error-correcting codes included in various modern communication standards, e.g., 5G and WLAN, due to their outstanding error correction performance with iterative message-passing decoding, typically known as belief propagation (BP) decoding. In particular, close-to-capacity performance is reported when decoding LDPC codes using the sum-product algorithm (SPA). However, the SPA requires the evaluation of computationally intensive check node updates~\cite{KLF2001}. Therefore, complexity-reduced variants like normalized min-sum (NMS)~\cite{AVAT2014} or differential decoding with binary message passing (DD-BMP)~\cite{MBH2009} have been introduced.

It is known that introducing memory to decoding via successive relaxation (SR) enhances its performance for the SPA and the min-sum (MS) algorithm~\cite{XTB2008}.
As pointed out in~\cite{JB2012}, DD-BMP is also related to SR due to the memory of the decoder. 
Hence, decoding with memory can reduce the gap between complexity-reduced BP and plain SPA. Furthermore, SPA with SR can outperform classical SPA.
In addition, a threshold to suppress small values can also help to improve MS decoding, as the offset MS decoder shows \cite{CDEFX2005}.

A system implementing the BP algorithm with SNNs has the potential to provide energy-efficient decoding.
Therefore, BP was implemented with SNNs in~\cite{SMD09} by connecting multiple SNNs and exchanging messages via spikes.
The deviation of SNN-based BP messages compared to the correct messages was measured for a small graph with six nodes. 
However, no practical channel code was evaluated in~\cite{SMD09}.

In this letter, we propose a novel BP decoder that incorporates SNNs.
In contrast to~\cite{SMD09}, the novel system uses simple SNNs rather than liquid and readout pools.
Furthermore, we rely on real-valued/graded spikes, compared to the encoding to a pulse rate of~\cite{SMD09}.
We optimized SNNs to replace the complex SPA check node update with energy-efficient SNN operations. 
Two decoding variants are introduced. The simplified variant requires SNNs consisting of only a single neuron. 
While both outperform NMS and DD-BMP decoding in Monte Carlo simulations, in high signal-to-noise ratio (SNR) regimes, the proposed decoder outperforms SPA, and the simplified variant achieves SPA-like performance.
To the best of our knowledge, this is the first work that uses SNNs to improve the energy efficiency of BP for channel decoding of practically relevant channel codes. 
\RevIII{
The implementation of the proposed SNN-based BP decoder and additional results are available at~\cite{github-code}.} \textcolor{white}{https://www.youtube.com/watch?v=GFq6wH5JR2A}\vspace*{-6mm}

\section{Spiking Neural Networks}
\begin{figure}
    \centering
    \resizebox{0.9\columnwidth}{!}{%
        \input{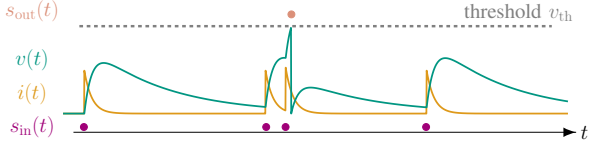}
    }
    \caption{Exemplary SNN dynamics. Input spikes $s_\mathrm{in}(t)$ (illustrated by dots) generate a synaptic current $i(t)$, which charges the membrane potential $v(t)$. If $v(t)$ exceeds the firing threshold $v_\mathrm{th}$, an output spike $s_\mathrm{out}(t)$ is generated.}
    \label{fig:lif_model}
    \vspace*{-0.5cm}
\end{figure}

\vspace*{-0.1cm}
\label{sec:Spiking Neural Networks}
SNNs consist of multiple layers of spiking neurons.
Like classical neural networks, the layers are connected by linear layers.
Spiking neurons can be interconnected within a layer, resulting in recurrent connections. 
Compared to classical neural networks, SNNs exhibit two significant differences:
First, they exchange information in short pulses, so-called \textit{spikes}. 
Second, spiking neurons are state-dependent and exhibit temporal dynamics, mimicking the biological neuron~\cite{NMZ2019}.
The \textit{membrane potential} $v(t)$ of a spiking neuron describes its state. 
An incoming spike induces a \textit{synaptic current}~$i(t)$, which charges~$v(t)$.
Concurrently,~$v(t)$ decreases over time. 
If~$v(t)$ exceeds the threshold potential~$v_\mathrm{th}$ of the neuron, an output spike is generated and the membrane potential is reset to the \textit{resting potential} $v_\mathrm{r}$.
The leaky integrate-and-fire (LIF) model is a simple yet biologically plausible and, therefore, common spiking neuron model.  
Its dynamics in differential form are~\cite{NMZ2019}
\vspace*{-1mm}
\begin{align}
    \frac{\mathrm{d}i(t)}{\mathrm{d}t}  &=-\frac{i(t)}{\tau_\mathrm{s}} + \sum_j w_j s_{\mathrm{in},j}(t)
    \label{eq:syn}, \\[-.1em]
    \frac{\mathrm{d}v(t)}{\mathrm{d}t}  &=-\frac{\left(\left(v(t)-v_\mathrm{r}\right)+i(t)\right)}{\tau_\mathrm{m}}  + s_\mathrm{out}(t)\left( v_\mathrm{r} - v_\mathrm{th} \right) .
    \label{eq:volt} 
\end{align}
Hereby, $\tau_\mathrm{s} \in \mathbb{R}^+$ denotes the time constant of the synaptic current, ${w_j~\in~\mathbb{R}}$ the weight applied to the $j$-th incoming sequence of spikes ${s_{\mathrm{in},j}(t) \in \mathbb{R}}$, $\tau_\mathrm{m} \in \mathbb{R}^+$ the time constant of the membrane potential, and ${s_\mathrm{out}(t)\in\{0,1\}}$ the output spike sequence generated by $s_\mathrm{out}(t)=\Theta(v(t)-v_\mathrm{th})$, where $\Theta(\cdot)$ denotes the Heaviside function. 
The input spike sequence~$s_{\mathrm{in},j}(t)$ can be either sent from downstream layers (feed-forward) or spiking neurons within the same layer (recurrent).
Fig.~\ref{fig:lif_model} provides an example of the dynamics of a LIF neuron.
The leaky integrate (LI) neuron is obtained by deactivating the spiking functionality. 
Hence, LI neurons act as memory with leakage.
For simulation and training, we utilize the PyTorch-based framework \textit{Norse} \cite{Norse}.

\vspace*{-0.3cm}

\section{Belief Propagation Decoding}
\label{sec:Belief Propagation Decoder}
\vspace*{-0.1cm}
We consider the decoding of binary $(N,k)$ LDPC codes, which map
$k$ data bits onto $N$ code bits.
LDPC codes are defined as the null space of a sparse parity-check matrix (PCM) $\bm{H}\in \mathbb{F}_2^{M\times N}$.
BP decoding is conducted on the Tanner graph, 
a bipartite graph associated with the respective PCM~\cite[pp.~51-59]{RU2008}. 
A Tanner graph consists of two types of nodes.
First, the check nodes (CNs) $\mathsf{c}_j$, with $j\in\{1,2,\ldots,M\}$, which correspond to the single parity checks, i.e., the rows of the PCM.  
The second type of nodes are variable nodes (VNs) $\mathsf{v}_i$, with $i\in\{1,2,\ldots,N\}$, where each node is associated with a code bit, i.e., a column of the PCM.
An edge between a CN $\mathsf{c}_j$ and a VN $\mathsf{v}_i$ in the Tanner graph exists iff $H_{j,i}=1$.
A regular ($d_\mathrm{v}$, $d_\mathrm{c}$) LDPC code possesses $d_\mathrm{c}$ non-zero entries per row and $d_\mathrm{v}$ non-zero entries per column in its PCM.
The check node degree and variable node degree are denoted by $d_\mathrm{c}$ and $d_\mathrm{v}$, respectively.

Assume that a random codeword $\bm{x}$ is binary phase shift keying (BPSK) modulated and transmitted over a binary-input additive white Gaussian noise (AWGN) channel, i.e., ${y_i=(-1)^{x_i}+n_i}$, with $n_i\sim \mathcal{N}(0,\sigma^2)$,

The bit-wise log-likelihood ratio (LLR) $L_i$ at the channel output w.r.t the $i$-th code bit is defined as \vspace*{-1mm}
\begin{align}
    L_i = \log\left( \frac{P(Y_i=y_i|X_i=0)}{P(Y_i=y_i|X_i=1)} \right).
\end{align}
Next, we consider the SPA, in which messages are LLRs that are iteratively updated in the nodes and passed along the edges of the Tanner graph.
A BP iteration with a flooding schedule includes the parallel updating of all CN messages and then all VN messages.
The variable-to-check-node messages are initialized with $L_{i\rightarrow j}^{[\mathrm{v}]}=L_i$, where $L_{i\rightarrow j}^{[\mathrm{v}]}$ is the variable-to-check-node message sent from  $\mathsf{v}_i$ to $\mathsf{c}_j$.
First, the check-to-variable-node messages $L_{i\leftarrow j}^{[\mathrm{c}]}$
are evaluated. To this end, we simplify the update equation \cite[(2.17)]{RU2008} by splitting it into the absolute value $\alpha_{i\leftarrow j}^{[\mathrm{c}]}=\left\vert L_{i\leftarrow j}^{[\mathrm{c}]}\right\vert$ and the sign~$\beta_{i\leftarrow j}^{[\mathrm{c}]}=\mathrm{sign} \left(L_{i\leftarrow j}^{[\mathrm{c}]}\right)$, i.e., \vspace*{-1mm}
\begin{align}
    L_{i\leftarrow j}^{[\mathrm{c}]} &= \alpha_{i\leftarrow j}^{[\mathrm{c}]} ~ \beta_{i\leftarrow j}^{[\mathrm{c}]}
                                ,\quad \forall i \in \mathcal{M}(j)\label{eq:cn_update} , \\
        \alpha_{i\leftarrow j}^{[\mathrm{c}]} &=  2\cdot \textrm{tanh}^{-1}
                                    \left( 
                                        \prod_{i' \in \mathcal{M}(j)\setminus \{i\}}
                                        \textrm{tanh}\left(\frac{|L_{i'\rightarrow j}^{[\mathrm{v}]}|}{2}\right) 
                                    \right)\label{eq:cn_update_alpha},\\
    \beta_{i\leftarrow j}^{[\mathrm{c}]} &= \prod_{i' \in \mathcal{M}(j)\setminus \{i\}} 
                                    \textrm{sign}\left( L_{i'\rightarrow j}^{[\mathrm{v}]} \right) \label{eq:cn_update_beta}  ,
\end{align}
where $\mathcal{M}(j)=\{i:H_{j,i}=1\}$ is the set of indices of VNs connected to CN $j$. 

Next, we introduce the VN update calculating the variable-to-check-node message sent from $\mathsf{v}_i$ to $\mathsf{c}_j$ using
\begin{align}
    L_{i\rightarrow j}^{[\mathrm{v}]} &= L_i + 
    \hspace*{-3mm}
    \sum_{j'\in \mathcal{N}(i)\setminus \{j\}} L_{i\leftarrow j'}^{[\mathrm{c}]} ,\quad \forall j \in \mathcal{N}(i)\,
    \label{eq:vn_update},
\end{align}
where $\mathcal{N}(i)=\{j:H_{j,i}=1\}$ is the set of indices of neighboring CNs of $\mathsf{v}_i$.

Note that both the CN and the VN update follow the extrinsic principle, i.e., when updating the message to VN $\mathsf{v}_i$ (CN $\mathsf{c}_j$), the message originating from
$\mathsf{v}_i$ (CN $\mathsf{c}_j$) is excluded from the update equation.
After reaching a predefined maximum number of iterations, the bit-wise output LLRs $\tilde{L}_i$ are calculated using $\tilde{L}_{i} = L_i + \sum_{j\in \mathcal{N}(i)} L_{i\leftarrow j}^{[\mathrm{c}]}$.
Finally, the decoded bits $\hat{b}_i$ are obtained by applying a hard decision to $\tilde{L}_i$, where $\hat{b}_i=0$ if $\tilde{L}_i>0$ and $\hat{b}_i=1$ otherwise.
For a more detailed explanation of message-passing algorithms, the interested reader is referred to~\cite[Ch.~2]{RU2008}.

\vspace*{-0.1cm}
\section{Spiking Neural Belief Propagation Decoding}\label{sec:SNNBP}

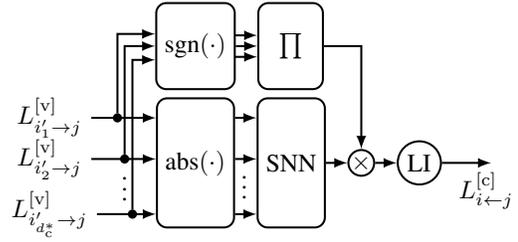
\begin{figure}[!t]
    \vspace*{-0.7cm}
    \begin{center}
    \resizebox{0.4\textwidth}{!}{
    \begin{tikzpicture}[>=latex,thick]
    \def\ydist{0.5cm}
    \def\xdist{0.3cm}
    \def\h{1.2cm}
    \node[] (L2) {$L_{i_2'\rightarrow j}^{[\mathrm{v}]}$};
    \node[above=-0.25cm of L2] (L1) {$L_{i_1'\rightarrow j}^{[\mathrm{v}]}$};
    \node[below=-0.1cm of L2] (L3) {$L_{i_{d^\ast_\mathrm{c}}'\rightarrow j}^{[\mathrm{v}]}$};

    \node[above right=-1.38cm and 3*\xdist of L2,draw,rectangle,rounded corners, minimum height=1.5*\h] (abs) {abs$(\cdot)$};
    \node[right=\xdist of abs,draw,rectangle, rounded corners,minimum height=1.5*\h,minimum width=0.9cm] (snn) {SNN};
    \node[right=\xdist of snn,draw,circle, inner sep=0pt,radius=0.1cm] (m1) {$\times$};

    \draw[->] (L1) to (\tikztostart -| abs.west);
    \draw[->] (L2) to (\tikztostart -| abs.west);
    \draw[->] (L3) to (\tikztostart -| abs.west);

    \draw[->] (L1 -| abs.east) to (\tikztostart -| snn.west);
    \draw[->] (L2 -| abs.east) to (\tikztostart -| snn.west);
    \draw[->] (L3 -| abs.east) to (\tikztostart -| snn.west);

    \draw[->] (snn) -- (m1);

    \node[right=\xdist of L1,draw,circle,inner sep=1, fill=black] (b1) {};
    \node[right=\xdist+0.1cm of L2,draw,circle,inner sep=1, fill=black] (b2) {};
    \node[right=\xdist+0.13cm of L3,draw,circle,inner sep=1, fill=black] (b3) {};
    \node[below=-0.15cm of b2] () {$\vdots$};
    \node[below right =-0.15cm and 1.45cm of b2] () {$\vdots$};

    \node[above=\ydist*0.2 of abs,draw, rectangle, rounded corners,minimum height=\h] (sgn) {sgn$(\cdot)$};
    \node[right=\xdist of sgn,draw,rectangle,rounded corners,,minimum height=\h,minimum width=0.9cm] (product) {$\prod$};

    \draw[->] (b1) -- +(0cm,1.17cm) to (\tikztostart -|sgn.west);
    \draw[->] (b2) |- (sgn);
    \draw[->] (b3) -- +(0cm,2.15cm) to (\tikztostart -|sgn.west);

    \draw[->] (sgn)++(0.55cm,0.15cm) to (\tikztostart -| product.west);
    \draw[->] (sgn) to (\tikztostart -| product.west);
    \draw[->] (sgn)++(0.55cm,-0.15cm) to (\tikztostart -| product.west);

    \draw[->] (product) -| (m1);

    \node[right=\xdist of m1,draw,circle, inner sep=2pt] (LI) {LI};
    \draw[->] (m1) -- (LI);

    \draw[->] (LI) -- ++(1cm,0cm) node[pos=0.9,below] () {$L_{i \leftarrow j}^{[\mathrm{c}]}$};

\end{tikzpicture}
    }
    \caption{Setup of an \Systemname~SCNU calculating the message $L_{i\leftarrow j}^{[\mathrm{c}]}$ based on the messages $L^{[\mathrm{v}]}_{i'\rightarrow j},\,i'\in \mathcal{M}(j)\setminus\{i\}=:\{i_1,\ldots,i_{d^\ast_\mathrm{c}}\}$ with ${d^\ast_\mathrm{c}=d_{\mathrm{c}}-1}$. The upper and lower branches realize~\eqref{eq:cn_update_alpha} and~\eqref{eq:cn_update_beta}, respectively. 
    The LI neuron implements the memory.}
    \label{fig:SPC_SRNCTMS}
    \end{center}
    \vspace*{-.8cm}
\end{figure}

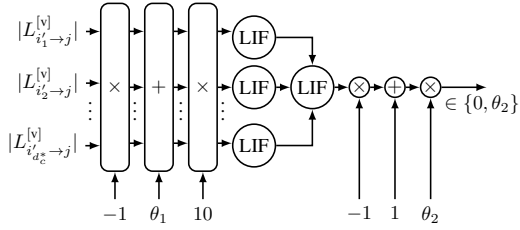
\begin{figure}[!t]
    \vspace*{-0.4cm}
    \begin{center}
    \resizebox{0.4\textwidth}{!}{
    \begin{tikzpicture}[>=latex,thick]
    \def\ydist{0.5cm}
    \def\xdist{0.25cm}
    \def\h{1.2cm}
    \node[] (L2) {$|L_{i_2'\rightarrow j}^{[\textrm{v}]}|$};
    \node[above= 0.1cm of L2] (L1) {$|L_{i_1'\rightarrow j}^{[\textrm{v}]}|$};
    \node[below left= 0.25cm and -1.35cm of L2] (L3) {$|L_{i_{d^\ast_c}'\rightarrow j}^{[\textrm{v}]}|$};

    \node[above right=-2.0cm and 1*\xdist of L2,draw,rectangle,rounded corners, minimum height=2.5*\h] (mul1) {$\times$};
    \draw[->] (mul1)++(0cm,-2.0cm) -- (mul1.south);
    \node[below=0.5cm of mul1] () {$-1$};

    \node[right=\xdist of mul1,draw,rectangle,rounded corners, minimum height=2.5*\h] (add1) {$+$};
    \draw[->] (add1)++(0cm,-2.0cm) -- (add1.south);
    \node[below=0.48cm of add1] () {$\theta_1$};

    \node[right=\xdist of add1,draw,rectangle,rounded corners, minimum height=2.5*\h] (mul2) {$\times$};
    \draw[->] (mul2)++(0cm,-2.0cm) -- (mul2.south);
    \node[below=0.5cm of mul2] () {$10$};
    
    \node[right=\xdist of mul2 ,draw,circle, inner sep=2pt,radius=2.0cm] (LIF2) {LIF};
    \node[above= 0.1cm of LIF2 ,draw,circle, inner sep=2pt,radius=2.0cm] (LIF1) {LIF};
    \node[below= 0.25cm of LIF2,draw,circle, inner sep=2pt,radius=2.0cm] (LIF3) {LIF};

    \node[right=\xdist of LIF2 ,draw,circle, inner sep=2pt,radius=2.0cm] (LIFOR) {LIF};
    
    \node[right=\xdist of LIFOR,draw,circle, inner sep=0pt,radius=2.0cm] (mul3) {$\times$};
    \draw[->] (mul3)++(0cm,-2.0cm) -- (mul3.south);
    \node[below=1.8cm of mul3] () {$-1$};
    
    \node[right=\xdist of mul3,draw,circle, inner sep=0pt,radius=2.0cm] (add3) {$+$};
    \draw[->] (add3)++(0cm,-2.0cm) -- (add3.south);
    \node[below=1.8cm of add3] () {$1$};
    
    \node[right=\xdist of add3,draw,circle, inner sep=0pt,radius=2.0cm] (mul4) {$\times$};
    \draw[->] (mul4)++(0cm,-2.0cm) -- (mul4.south);
    \node[below=1.8cm of mul4] () {$\theta_2$};

    \draw[->] (L1) to (\tikztostart -| mul1.west);
    \draw[->] (L2) to (\tikztostart -| mul1.west);
    \draw[->] (L3) to (\tikztostart -| mul1.west);

    \draw[->] (mul1.east) to (add1.west);
    \draw[->] (mul1.east)++(0,1cm) -- ++(\xdist,0cm);
    \draw[->] (mul1.east)++(0,-1cm) -- ++(\xdist,0cm);

    \draw[->] (add1.east) to (mul2.west);
    \draw[->] (add1.east)++(0,1cm) -- ++(\xdist,0cm);
    \draw[->] (add1.east)++(0,-1cm) -- ++(\xdist,0cm);

    \draw[->] (mul2.east) to (LIF2.west);
    \draw[->] (mul2.east)++(0,1cm) -- ++(\xdist,0cm);
    \draw[->] (mul2.east)++(0,-1cm) -- ++(\xdist,0cm);

    \draw[->] (LIF1.east) -| (LIFOR.north);
    \draw[->] (LIF2.east) -- (LIFOR.west);
    \draw[->] (LIF3.east) -| (LIFOR.south);

    \draw[->] (LIFOR) -- (mul3);
    \draw[->] (mul3) -- (add3);
    \draw[->] (add3) -- (mul4);

    \draw[->] (mul4) -- ++(1.0cm,0cm) node[pos=0.9,below] () {$\in\{0,\theta_2\}$};

    \node[above right=-1.2cm and -0.1cm of L2] () {$\vdots$};
    \node[above right=-1.2cm and 0.7cm of L2] () {$\vdots$};
    \node[above right=-1.2cm and 1.52cm of L2] () {$\vdots$};
    \node[above right=-1.2cm and 2.27cm of L2] () {$\vdots$};
    
\end{tikzpicture}
    }
    \caption{Setup of the SNN block in Fig.~\ref{fig:SPC_SRNCTMS} implementing~\eqref{eq:elena} using $d_\mathrm{c}$ LIF-neurons.
    If an input value is below $\theta_1$, the connected neuron is charged to emit a spike. The upstream neuron combines all incoming spikes and forwards them, leading to a zero output. 
    If all input values are larger than $\theta_1$, no neuron is charged, leading to $\theta_2$ as output.}
    \label{fig:SNN}
    \end{center}
    \vspace*{-0.7cm}
\end{figure}

Equation~\eqref{eq:cn_update_alpha} constitutes a significant part of the complexity of the SPA~\cite{KLF2001}. 
Therefore, an efficient replacement of the CN update using, e.g., SNNs, can improve the energy efficiency of the decoder. 
Hence, we propose a novel decoder named Enlarge-Likelihood-Each-Notable-Amplitude Spiking-Neural-Network decoder, referred to as \textit{\Systemname}.
\Systemname~uses a so-called SNN CN update (SCNU) substitute for integrating SNNs into the CN update equations~\eqref{eq:cn_update}-\eqref{eq:cn_update_beta}. \\
Fig.~\ref{fig:SPC_SRNCTMS} depicts the setup of an SCNU calculating the message $L^{[\mathrm{c}]}_{i\leftarrow j}$ given the messages $L^{[\mathrm{v}]}_{i'\rightarrow j},\,i'\in \mathcal{M}(j)\setminus\{i\}$.
The incoming $L_{i'\rightarrow j}^{[\mathrm{v}]}$ are split into their sign and absolute value; both are processed, combined, and afterward integrated over time by an LI neuron, which acts as memory. 
Therefore, $\tau_m$ and $\tau_s$ of the LI neuron control the behavior of the SCNU memory.
The upper branch performs~(\ref{eq:cn_update_beta}), which can be implemented using XOR operations. 
The lower branch realizes~(\ref{eq:cn_update_alpha}), where the computationally complex operations of~(\ref{eq:cn_update_alpha}) are replaced by the SNN, shown in Fig.~\ref{fig:SNN}.
Inspired by the offset MS algorithm~\cite{CDEFX2005}, which approximates~(\ref{eq:cn_update_alpha}) by 
${\alpha_{i\leftarrow j}^{[\mathrm{c}]}\approx \max \left\{\mathrm{min}_{i' \in \mathcal{M}(j)\setminus \{i\}}\, \left\vert L_{i'\rightarrow j}^{[\mathrm{v}]} \right\vert - \theta_1 ,0  \right\}}$, we further simplify the approach by \vspace*{-1mm}
\begin{align} 
    \alpha_{i\leftarrow j}^{[\mathrm{c}]}\approx \begin{cases}
        \theta_2 &\text{if}\; \min_{i' \in \mathcal{M}(j)\setminus \{i\}} \, \left\vert L_{i'\rightarrow j}^{[\mathrm{v}]} \right\vert > \theta_1\, , \\
        0 &\text{otherwise}\, .
    \end{cases} \; .
\label{eq:elena}
\end{align}
Like the offset MS algorithm,~(\ref{eq:elena}) returns zero if the input values are below the threshold $\theta_1\in\mathbb{R}^{+}$.
In contrast to the offset MS algorithm, which returns the biased minimum value,~(\ref{eq:elena}) returns a fixed value $\theta_2\in\mathbb{R}^{+}$ if the threshold $\theta_1$ is exceeded.
\RevII{Hence, we approximate the term $\mathrm{min}_{i' \in \mathcal{M}(j)\setminus \{i\}}\, \vert L_{i'\rightarrow j}^{[\mathrm{v}]} \vert - \theta_1$ with $\theta_2$.}
Fig.~\ref{fig:SNN} visualizes the implementation of~(\ref{eq:elena}) using LIF spiking neurons.
\RevII{To implement \eqref{eq:elena}, we emit zero if at least one input value $\vert L_{i'\rightarrow j}^{[\mathrm{v}]} \vert$ is below the threshold $\theta_1$, and $\theta_2$ otherwise.
Therefore, the signs of all inputs $\vert L_{i'\rightarrow j}^{[\mathrm{v}]} \vert$ are inverted, the bias $\theta_1$ is added, the signal is amplified and integrated by LIF neurons.
If $(\theta_1 - \vert L_{i'\rightarrow j}^{[\mathrm{v}]}\vert) >0$, the connected LIF neuron is charged, an output spike is generated and passed to the combining LIF neuron.
If $(\theta_1 - \vert L_{i'\rightarrow j}^{[\mathrm{v}]} \vert) <0$, the connected LIF neurons are discharged, and no output spike is generated.
Hence, the multiplications, the adder, and the LIF neurons implement the inverted sign function; if a value is below $\theta_1$, a spike is emitted. 
The following combining LIF neuron fires if any of the previous neurons emit a spike. 
If the combining LIF neuron emits a spike, the output turns zero.
Hence, if the minimum absolute of the input values is above $\theta_1$, $\theta_2$ is emitted.
If at least one value is below $\theta_1$, zero is emitted.}
\\

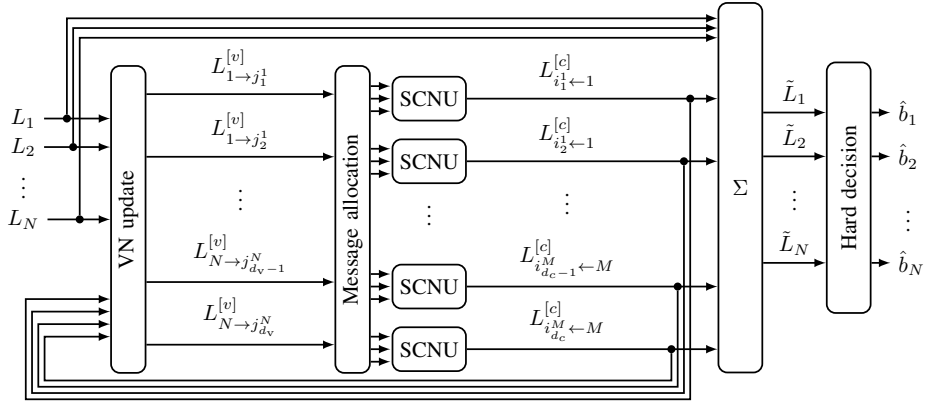
\begin{figure*}[!t]
    \vspace*{-.7cm}
    \begin{center}
    \resizebox{0.7\textwidth}{!}{
    \begin{tikzpicture}[>=latex, thick]
\def\ydist{0.5cm}
        \def\xdist{1cm}
    
        \node [] (L1) {$L_1$};
        \node[below = -0.1cm of L1] (L2) {$L_2$};
        \node [below of =  L1] (h) {$\vdots$};
        \node[below = 1.2*\ydist of L2] (LN) {$L_N$};
    
        \node[right=1*\xdist of LN, draw, rectangle, rounded corners, align=center,minimum height = 4.9cm] (VN_up) {\rotatebox{90}{VN update}};  
        \draw[->] (L1) to (\tikztostart -| VN_up.west);
        \draw[->] (L2) to (\tikztostart -| VN_up.west);
        \draw[->] (LN) to (\tikztostart -| VN_up.west);
    
        \node[right=3*\xdist of VN_up, draw, rectangle, rounded corners, minimum height = 4.9cm] (dist) {\rotatebox{90}{Message allocation}};
        \draw[->] (VN_up.east)++(0cm,2cm) -- +(3*\xdist,0cm) node[midway, above] {$L_{1 \rightarrow j_1^1}^{[v]}$ };        
        \draw[->] (VN_up.east)++(0cm,1cm) -- +(3*\xdist,0cm) node[midway, above] (h1) {$L_{1 \rightarrow j_2^1}^{[v]}$};
        \draw[->] (VN_up.east)++(0cm,-1cm) -- +(3*\xdist,0cm) node[midway, above] ()
        {$L_{N \rightarrow j_{d_\mathrm{v}-1}^N}^{[v]}$};
        \draw[->] (VN_up.east)++(0cm,-2cm) -- +(3*\xdist,0cm) node[midway, above] ()
        {$L_{N \rightarrow j_{d_\mathrm{v}}^N}^{[v]}$};
        \node[below of=h1] () {$\vdots$};
        
        \def\h{0.7cm}
        \node [above right = -0.9cm and \xdist/3 of dist, draw, rectangle, rounded corners, minimum height = \h] (SCNU1) {SCNU};
        \node [above right = -1.9cm and \xdist/3 of dist, draw, rectangle, rounded corners, minimum height = \h] (SCNU2) {SCNU};
        \node [above right = -3.9cm and \xdist/3 of dist, draw, rectangle, rounded corners, minimum height = \h] (SCNU3) {SCNU};
        \node [above right = -4.9cm and \xdist/3 of dist, draw, rectangle, rounded corners, minimum height = \h] (SCNUM) {SCNU};
        \node [below=of SCNU1] (h2) {$\vdots$};

        \def\yoff{0.2cm}
        \draw[<-] (SCNU1)+(-0.62cm,\yoff) to (\tikztostart -| dist.east);
        \draw[<-] (SCNU1)+(-0.62cm,0.0cm) to (\tikztostart -| dist.east);
        \draw[<-] (SCNU1)+(-0.62cm,-\yoff) to (\tikztostart -| dist.east);

        \draw[<-] (SCNU2)+(-0.62cm,\yoff) to (\tikztostart -| dist.east);
        \draw[<-] (SCNU2)+(-0.62cm,0.0cm) to (\tikztostart -| dist.east);
        \draw[<-] (SCNU2)+(-0.62cm,-\yoff) to (\tikztostart -| dist.east);

        \draw[<-] (SCNU3)+(-0.62cm,\yoff) to (\tikztostart -| dist.east);
        \draw[<-] (SCNU3)+(-0.62cm,0.0cm) to (\tikztostart -| dist.east);
        \draw[<-] (SCNU3)+(-0.62cm,-\yoff) to (\tikztostart -| dist.east);
        
        \draw[<-] (SCNUM)+(-0.62cm,\yoff) to (\tikztostart -| dist.east);
        \draw[<-] (SCNUM)+(-0.62cm,0.0cm) to (\tikztostart -| dist.east);
        \draw[<-] (SCNUM)+(-0.62cm,-\yoff) to (\tikztostart -| dist.east);

        \def\x{4cm}
        \node [above right=-1.75cm and \x of SCNU3, draw, rectangle, rounded corners, minimum height=5.9cm,minimum width =0.7cm] (sum) {$\Sigma$};
        \draw[->] (SCNU1) -- +(\x+0.6cm,0cm) node[pos=0.4, above] () 
        {$L_{i^1_1 \leftarrow 1}^{[c]}$}; 
        \draw[->] (SCNU2) -- +(\x+0.6cm,0cm) node[pos=0.4, above] (hc) {$L_{i^1_2 \leftarrow 1}^{[c]}$};
        \draw[->] (SCNU3) -- +(\x+0.6cm,0cm) node[pos=0.4, above] () {$L_{i^M_{d_c-1} \leftarrow M}^{[c]}$};
        \draw[->] (SCNUM) -- +(\x+0.6cm,0cm) node[pos=0.4, above] () {$L_{i^M_{d_c} \leftarrow M}^{[c]}$};
        \node[below of=hc] () {$\vdots$};

        \node [right=0.3cm of L1, draw, circle, inner sep =1pt, fill=black] (n1) {};
        \node [right = 0.4 of L2, draw, circle, inner sep =1pt, fill=black] (n2) {};
        \node [right=0.45cm of LN, draw, circle, inner sep =1pt, fill=black] (n3) {};
        \draw[->] (n1) -- +(0cm,1.6cm) to (\tikztostart -| sum.west);
        \draw[->] (n2) -- +(0cm,1.9cm) to (\tikztostart -| sum.west);
        \draw[->] (n3) -- +(0cm,2.9cm) to (\tikztostart -| sum.west);

        \node[right=\x-0.5cm of SCNU1, draw, fill = black, circle, inner sep =1pt] (m1) {};
        \node[right=\x-0.6cm of SCNU2, draw, fill = black, circle, inner sep =1pt] (m2) {};
        \node[right=\x-0.7cm of SCNU3,draw, fill = black,circle, inner sep=1pt] (m3) {};
        \node[right=\x-0.8cm of SCNUM, draw, fill = black, circle, inner sep =1pt] (m4) {};
        \draw[->] (m1) -- +(0cm,-4.8cm) -- +(-10.6cm,-4.8cm) -- +(-10.6cm,-3.2cm) to (\tikztostart -| VN_up.west);
        \draw[->] (m2) -- +(0cm,-3.7cm) -- +(-10.4cm,-3.7cm) -- +(-10.4cm,-2.4cm) to (\tikztostart -| VN_up.west);
        \draw[->] (m3) -- +(0cm,-1.6cm) -- +(-10.2cm,-1.6cm) -- +(-10.2cm,-0.6cm) to (\tikztostart -| VN_up.west);
        \draw[->] (m4) -- +(0cm,-0.5cm) -- +(-10cm,-0.5cm) -- +(-10cm,0.2cm) to (\tikztostart -| VN_up.west);

        \node [right=\xdist of sum, draw, rectangle, rounded corners, minimum height = 4cm, minimum width =0.7cm] (dec) {\rotatebox{90}{Hard decision}};
        \draw[->] (sum.east)++(0cm,1.2cm) -- +(\xdist,0cm) node [midway, above] (h3) {$\tilde{L}_1$};
        \draw[->] (sum.east)++(0cm,0.5cm) -- +(\xdist,0cm) node [midway, above] {$\tilde{L}_2$};
        \draw[->] (sum.east)++(0cm,-1.2cm) -- +(\xdist,0cm) node [midway, above] {$\tilde{L}_N$};
        \node[below = 0.9cm of h3] () {$\vdots$};

        \draw[->] (dec.east)++(0cm,1.2cm) -- +(0.3cm,0cm) node [right] (h4) {$\hat{b}_1$};
        \draw[->] (dec.east)++(0cm,0.5cm) -- +(0.3cm,0cm) node [right] {$\hat{b}_2$};
        \draw[->] (dec.east)++(0cm,-1.2cm) -- +(0.3cm,0cm) node [right] {$\hat{b}_N$};
        \node[below = 0.9cm of h4] () {$\vdots$};
\end{tikzpicture}
    }
    \caption{The architecture of the proposed decoder. The SCNU contains the SNN. 
    For $i\in \{1,\ldots,N\}$ and $j\in \{1,\ldots,M\}$, we use the notation $\mathcal{N}(i):=\{j_1^i,\ldots, j^i_{d_v}\}$ and $ \mathcal{M}(j):=\{i_1^j,\ldots, i^j_{d_v}\}$, respectively.
    \label{fig:SNN_BP_Structure}}
    \end{center}
    \vspace*{-0.7cm}
\end{figure*}

Fig.~\ref{fig:SNN_BP_Structure} depicts the overall decoder architecture, integrating the SCNUs. 
All SCNUs share the same parameters.
First, for initialization, all $L_{i\leftarrow j}^{[\mathrm{c}]}$ are set to zero, and the VNs are updated according to~\eqref{eq:vn_update}.
Next, we start the iterative decoding process. 
The update of each CN consists in allocating the variable-to-check-node messages $L_{i\rightarrow j}^{[\mathrm{v}]}$ to the respective SCNU. Assuming a regular LDPC code, there exist $M\cdot d_{\mathrm{c}}$ SCNUs, each performing an extrinsic CN update. Hence, in the allocation, all variable-to-check node messages $L_{i'\rightarrow j}^{[\mathrm{v}]}$ with ${i'\in \mathcal{M}(j)\setminus\{i\}}$ participating in the update of $L_{i\leftarrow j}^{[\mathrm{c}]}$ are routed to the respective SCNU, which implements \eqref{eq:cn_update}. 
Afterward, the VNs are updated, and the process is iteratively repeated until the hard decision maps the output LLRs to binary values.
After that, the membrane potentials and synaptic currents of all neurons are reset to zero. 
The memory is, therefore, also reset.
It is important to note that the decoder uses a flooding schedule, i.e., all node updates are performed in parallel.

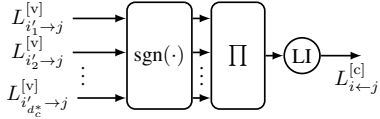
\begin{figure}[!t]
    \begin{center}
    \resizebox{0.3\textwidth}{!}{
    \begin{tikzpicture}[>=latex,thick]
    \def\ydist{0.5cm}
    \def\xdist{0.3cm}
    \def\h{1.2cm}
    \node[] (L2) {$L_{i_2'\rightarrow j}^{[\mathrm{v}]}$};
    \node[above=-0.25cm of L2] (L1) {$L_{i_1'\rightarrow j}^{[\mathrm{v}]}$};
    \node[below=-0.1cm of L2] (L3) {$L_{i_{d^\ast_c}'\rightarrow j}^{[\mathrm{v}]}$};

    \node[above right=-1.38cm and 3*\xdist of L2,draw,rectangle,rounded corners, minimum height=1.5*\h] (sgn) {sgn$(\cdot)$};
    \node[right=\xdist of sgn,draw,rectangle, rounded corners,minimum height=1.5*\h,minimum width=0.9cm] (prod) {$\prod$};
    \node[right=\xdist of prod,draw,circle, inner sep=2pt,radius=2.0cm] (LI) {LI};

    \draw[->] (L1) to (\tikztostart -| sgn.west);
    \draw[->] (L2) to (\tikztostart -| sgn.west);
    \draw[->] (L3) to (\tikztostart -| sgn.west);

    \draw[->] (L1 -| sgn.east) to (\tikztostart -| prod.west);
    \draw[->] (L2 -| sgn.east) to (\tikztostart -| prod.west);
    \draw[->] (L3 -| sgn.east) to (\tikztostart -| prod.west);

    \draw[->] (prod) -- (LI);

    \node[above right=-1.1cm and -0.0cm of L2] () {$\vdots$};
    \node[above right=-1.1cm and 2.0cm of L2] () {$\vdots$};
    
    \draw[->] (LI) -- ++(1cm,0cm) node[pos=0.9,below] () {$L_{i \leftarrow j}^{[\mathrm{c}]}$};
    
\end{tikzpicture}
    }
    \caption{Simplified \Systemname\Discriminator~SCNU. The multiplication of the signs and the LI neuron act similarly to DD-BMP.}
    \label{fig:SPC_DDBMP}
    \end{center}
    \vspace*{-0.7cm}
\end{figure}

Each SCNU of the \Systemname~decoder uses $d_\mathrm{c}$ LIF spiking neurons and a single LI neuron. 
We furthermore propose a simplified version of \Systemname, called \textit{\Systemname\Discriminator}. 
Inspired by the DD-BMP, which only takes the sign of each $L^{[\mathrm{v}]}_{i\rightarrow j}$ into account, the number of neurons per SCNU can be reduced to a single LI neuron by omitting the lower path of~Fig.~\ref{fig:SPC_SRNCTMS}. 
Its simplified SCNU can be seen in Fig.~\ref{fig:SPC_DDBMP}.

\vspace*{-0.3cm}
\section{Results and Discussion}
\vspace*{-2mm}
\label{sec:Results}
We implemented and evaluated the two proposed decoders for two different finite-geometry (FG) LDPC codes, the ($273$,$191$) and the ($1023$,$781$) code~\cite{KLF2001}, in the following called \Cs~and \Cb, respectively. Both FG LDPC codes are regular overcomplete codes, \Cs~has $d_\textrm{c}=17,\,d_\textrm{v}=17$, whereas \Cb~has $d_\textrm{c}=32,\,d_\textrm{v}=32$. \RevI{FG LDPC codes together with BP decoding are known to typically possess very low error floors, which are of interest for, e.g., storage and wireline applications. 
Currently, due to the highly irregular structure of the 5G LDPC code and the presence of low-degree VNs, \Systemname~does not yet achieve competitive performance. Adapting \Systemname~ to perform better with such codes is part of our ongoing investigation.}

For $y_i$, which is the received value corresponding to the $i$-th code bit of a codeword, 
the bit-wise LLRs are $L_i = y_i\cdot L_\mathrm{c}$, where $L_\mathrm{c}$ is the channel reliability parameter. 
Usually, for an AWGN channel, $L_\mathrm{c}=4\frac{E_\mathrm{s}}{N_0}=\frac{2}{\sigma^2}$ and therefore needs to be adapted to the channel SNR. 
However, depending on the code, we optimize both decoders at a fixed $\frac{E_\mathrm{s}}{N_0}$ corresponding to a BER of approximately $10^{-4}$.
For evaluation, the value of $L_\mathrm{c}$ is maintained, regardless of the channel SNR.
\RevII{
To compare against the fixed $L_\mathrm{c}$, the~\Systemname-$L_\mathrm{c}$ decoder adjusts $L_\mathrm{c}$ to match the channel SNR.
}
We choose $\frac{E_\mathrm{s}}{N_0}=3.5\,\mathrm{dB}$ for \Cs~and $\frac{E_\mathrm{s}}{N_0}=3.4\,\mathrm{dB}$ for \Cb.
For 20 decoding iterations, the proposed decoders are compared to the following benchmark decoders: SPA, DD-BMP, MS, NMS, and NMS with SR decoding (SR-NMS). \RevII{Further increasing the decoding iterations did not yield a significant improvement in the decoding performance.}
For all benchmark decoders, $L_\mathrm{c}$ matches the actual $E_\mathrm{s}/N_0$.

\vspace*{-0.4cm}
\subsection{Parameters}
\vspace*{-0.1cm}
For all LIF spiking neurons of the \Systemname~decoder, we fix $\tau_\mathrm{m}=\SI{1}{\milli\second}$, $\tau_\mathrm{s}=\SI{1}{\milli\second}$ and $v_\mathrm{th}=1$.
Furthermore, we fix $\tau_\mathrm{s}=\SI{1}{\milli\second}$ of the LI neurons.
Hence, the remaining parameters subject to optimization are $\tau_\mathrm{m}$ of the LI neurons as well as $\theta_1$ and $\theta_2$.
For the \Systemname\Discriminator~decoder, $\tau_\mathrm{s}=\SI{1}{\milli\second}$ of the LI neurons is also fixed.
Hence, the set of tuneable parameters is reduced to $\tau_\mathrm{m}$ of the LI neuron.

\begin{figure*}[!t]
\vspace*{-6mm}
    \centering  
    \minipage{0.32\textwidth}
        \resizebox{!}{.225\textheight}{\pgfplotsset{
layers/my layer set/.define layer set={
background,
main,
up
}{
 },
    set layers=my layer set,
}

\begin{tikzpicture}[spy using outlines={rectangle, magnification=2.7, size=1cm, connect spies}]
    \def\lwidth{1.5}
    \def\opac{50}
    \def\marksz{2pt}

    \begin{axis}[
        width=3in,
        height=3in,
        ymode = log,
        xscale=1,
        xlabel style = {align=center},
        xlabel=$\tau_\mathrm{m}~\textrm{in ms}$,
        ylabel=BER,
        y label style={at={(axis description cs:0.03,0.4)},anchor=west},
        xtick = {1,2,3,4,5,6,7,8,9,10},
        grid=major,
        legend cell align={left},
        legend style={
            at={(0.05,0.99)},
            anchor=north west,
            fill opacity = 0.8,
            draw opacity = 1, 
            text opacity = 1,
        },
        xmin=1,xmax=10,
        ymin=5e-5,ymax=1e-2,
        axis line style=thick,
        tick label style={/pgf/number format/fixed},
        ]

        \addplot[dashed,color=KITcyanblue, line width=\lwidth] table [x=taumem, y=SNN1_273_191]{./figures/TauMemSweep.txt};
        \addplot[dashed,color=KITorange, line width=\lwidth] table [x=taumem, y=SNN1_1023_781]{./figures/TauMemSweep.txt};
        \addplot[color=KITcyanblue, line width=\lwidth] table [x=taumem, y=SNN2_273_191]{./figures/TauMemSweep.txt};
        \addplot[color=KITorange, line width=\lwidth] table [x=taumem, y=SNN2_1023_781]{./figures/TauMemSweep.txt};

        \addlegendimage{} \addlegendentry{\Systemname\Discriminator~\Cs}
        \addlegendimage{} \addlegendentry{\Systemname\Discriminator~\Cb}
        \addlegendimage{} \addlegendentry{\Systemname~\Cs}
        \addlegendimage{} \addlegendentry{\Systemname~\Cb}

    \begin{pgfonlayer}{up}
        
    \end{pgfonlayer}
\end{axis}
\end{tikzpicture}}\\[-.8em]
        \footnotesize \hspace*{4mm}(a) \parbox{.8\textwidth}{ \textcolor{white}{A} \\ BER over $\tau_\mathrm{m}$ for the proposed decoders for both codes \RevII{with $\theta_1=\theta_2=1$}.}%
        \label{fig:epsilon_sweep}
    \endminipage
    \hspace{1mm}
    \minipage{0.32\textwidth}
    \vspace{0.3mm}
        \resizebox{.26347\textheight}{!}{\begin{tikzpicture}
    \begin{axis}[
        width=3in,
        height=3in,
        ymode = log,
        xlabel=$\theta_1$,
        ylabel=BER,
        y label style={at={(axis description cs:0.03,0.4)},anchor=west},
        grid=major,
        legend cell align={left},
        legend style={
            at={(0.05,0.99)},
            anchor=north west,
            fill opacity = 0.8,
            draw opacity = 1, 
            text opacity = 1,
        },
        xmin=0,xmax=3,
        ymin=5e-5,ymax=1e-3,
        axis line style=thick,
        tick label style={/pgf/number format/fixed},
        axis y line* = left,        %
        ]
        \addplot [color=KITcyanblue,solid,line width=1.5pt] table [x=thbias, y=BER]{./figures/Thbias_Sweep.txt}; %
        \addlegendentry[color=KITcyanblue,solid,line width=2.0pt]{\textcolor{black}{BER}};
    \end{axis}

    \begin{axis}[%
        width=3in,
        height=3in,
        xmin=0,
        xmax=3,
        ymin=0,
        ymax=2.0,
        ytick={   0, 0.5,  1.0, 1.5,   2.0},
        yticklabels ={    0$\%$, 0.5$\%$,  1$\%$, 1.5$\%$,  2$\%$},
        ylabel={Spike rate},
        y label style={at={(axis description cs:1.4,.483)},anchor=south},
        axis x line*=bottom,
        axis y line*=right,
        xtick={},
        xticklabels={},
        legend cell align={left},
        legend style={
            at={(0.95,0.99)},
            anchor=north east,
            fill opacity = 0.8,
            draw opacity = 1, 
            text opacity = 1,
        },
        ]
        \addplot [color=KITorange,solid,line width=1.5pt] table [x=thbias, y=REG]{./figures/Thbias_Sweep.txt};%
        \addlegendentry{Spike rate};
    \end{axis}
    
\end{tikzpicture}}\\[-0.8em]
        \footnotesize \hspace*{4mm}(b) \parbox{.8\textwidth}{ \textcolor{white}{A} \\ BER over $\theta_1$ for the \Systemname~decoder and \Cs~\RevII{with $\tau_\mathrm{m}=\SI{1}{\milli\second}$ and $\theta_2=1$}.}%
        \label{fig:bias_sweep}
    \endminipage
    \hspace{4mm}
    \minipage{0.32\textwidth}
    \vspace{.5mm}
        \resizebox{!}{.225\textheight}{\pgfplotsset{
layers/my layer set/.define layer set={
background,
main,
up
}{
 },
    set layers=my layer set,
}

\begin{tikzpicture}[spy using outlines={rectangle, magnification=2.7, size=1cm, connect spies}]
    \def\lwidth{1.5}
    \def\opac{50}
    \def\marksz{2pt}

    \def\spywidth{2.0cm}
    \def\spyheigth{1.5cm}

    \begin{axis}[
        width=3in,
        height=3in,
        ymode = log,
        xscale=1,
        xlabel style = {align=center},
        xlabel={$\theta_2$},
        ylabel=BER,
        y label style={at={(axis description cs:0.03,0.4)},anchor=west},
        xtick = {1,2,3,4,5,6,7,8,9,10},
        grid=major,
        legend cell align={left},
        legend style={
            at={(0.05,0.99)},
            anchor=north west,
            fill opacity = 0.8,
            draw opacity = 1, 
            text opacity = 1,
        },
        xmin=0,xmax=4,
        ymin=4e-5,ymax=1e-1,
        axis line style=thick,
        tick label style={/pgf/number format/fixed},
        ]

        \addplot[color=KITcyanblue, line width=\lwidth] table [x=amp, y=BER]{./figures/AmpSweep.txt};

    \begin{pgfonlayer}{up}
        
    \end{pgfonlayer}
\end{axis}
\end{tikzpicture}}\\[-0.8em]
        \footnotesize \hspace*{4mm}(c) \parbox{.8\textwidth}{ \textcolor{white}{A} \\ BER over $\theta_2$ for the \Systemname~decoder and \Cs~\RevII{with $\tau_\mathrm{m}=\SI{1}{\milli\second}$ and $\theta_1=2$}.}%
        \label{fig:amp_sweep}
    \endminipage
    \vspace*{-1mm}
    \caption{Impact of the parameters of the decoders on the performance. Both decoders are run for 20 iterations. \Cs~is run at~$\frac{E_\mathrm{s}}{N_0}=3.5\,\mathrm{dB}$ and \Cb~at~$\frac{E_\mathrm{s}}{N_0}=3.4\,\mathrm{dB}$. }
    \label{fig:sweep}
    \vspace{-5mm}
\end{figure*}
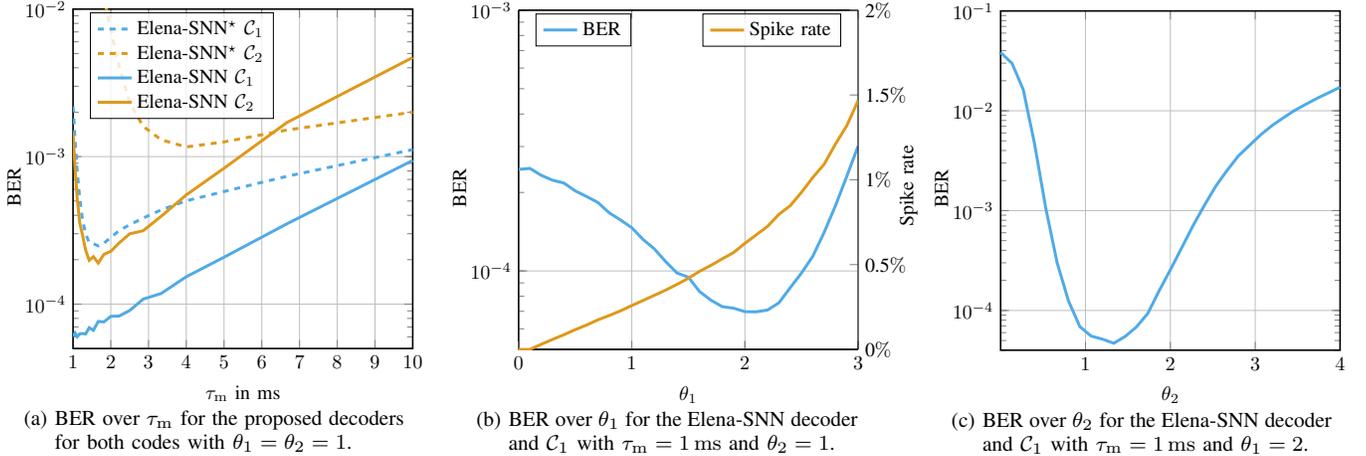

\RevII{
The membrane time constants $\tau_\mathrm{m}$ for the \Systemname~decoder and \Systemname\Discriminator~decoder are determined by a simple line search with $\theta_1=\theta_2=1$.
With the obtained $\tau_\mathrm{m}$, another line search yields $\theta_1$.
The obtained $\tau_\mathrm{m}$ and $\theta_1$ are fixed to determine $\theta_2$ in a final line search.
}
Fig.~\ref{fig:sweep}(a) shows the impact of $\tau_\mathrm{m}$ on the performance of both decoders for both codes. 
For \Systemname,~small $\tau_\mathrm{m}$ yield the best performance, reducing the averaging effect of the LI neuron.
\Systemname\Discriminator~performs best at larger $\tau_\mathrm{m}$, demonstrating the need for averaging by the LI neuron.
For \Systemname~and \Cs, Fig. \ref{fig:sweep}(b) shows the effect of the bias $\theta_1$ on the BER and the spike rate of the SNN, where the spike rate is the ratio of spikes generated by all LIF spiking neurons to possible spikes of the discrete SNN simulation.
Up to a value of $\theta_1\approx2$, an increase in $\theta_1$ leads to a decreasing BER, while the spike rate increases.
Further increasing $\theta_1$ yields both a higher BER and a higher spike rate.
For \Cb, a similar behavior was observed.
As Fig.~\ref{fig:sweep}(c) shows, the output amplitude $\theta_2$ significantly impacts the BER.
Again, for \Cb, a similar behavior was observed.
For the \Systemname~decoder we have chosen ($\tau_\mathrm{m}=\SI{1}{\milli\second}$, $\theta_1=2$, $\theta_2=1.4$) for \Cs~and ($\tau_\mathrm{m}=\SI{1.667}{\milli\second}$, $\theta_1=1.6$, $\theta_2=1$) for \Cb.
For the \Systemname\Discriminator~decoder we have chosen ${\tau_\mathrm{m}=\SI{1.639}{\milli\second}}$ for \Cs~and ${\tau_\mathrm{m}=\SI{4.167}{\milli\second}}$ for \Cb.

\vspace*{-0.35cm}
\subsection{Evaluation}
\begin{figure}[hbt!]
    \centering
    \begin{minipage}{1\textwidth}
        \resizebox{.45\textwidth}{!}{\pgfplotsset{
layers/my layer set/.define layer set={
background,
main,
up
}{
 },
    set layers=my layer set,
}

\begin{tikzpicture}[spy using outlines={rectangle, magnification=2.7, size=2cm, connect spies}]
    \def\lwidth{1}
    \def\opac{70}
    \def\marksz{1.5pt}
    \def\markszl{3pt}

    \def\spywidth{2.0cm}
    \def\spyheigth{1.5cm}

    \begin{axis}[
        ymode = log,
        xscale=1,
        xlabel style = {align=center},
        xlabel=$E_\mathrm{b}/N_0$ (dB),
        x label style={at={(axis description cs:0.5,0.02)},anchor=north},
        ylabel=BER,
        y label style={at={(axis description cs:0.00,0.4)},anchor=west},
        xtick = {0,1,2,3,4,5},
        xticklabels={$0$,$1$,$2$,$3$,$4$,$5$},
        grid=major,
        legend cell align={left},
        legend style={
            at={(0.02,0.01)},
            anchor=south west,
            fill opacity = 0.8,
            draw opacity = 1, 
            text opacity = 1,
        },
        xmin=0,xmax=5.0,
        ymin=1.0e-10,ymax=1e0,
        axis line style=thick,
        tick label style={/pgf/number format/fixed},
        legend image post style={scale=2},
        ]

        \coordinate (spypoint) at (axis cs:3.9,1.0e-5); %
        \coordinate (spyviewer) at (axis cs:4.1,5e-2); %
        \draw [fill=white] ($(spyviewer)+(1.2cm,0.7cm)$) rectangle ($(spyviewer)-(1.2cm,0.7cm)$);
        
        \spy[width=2.4cm,height=1.4cm, every spy on node/.append style={ultra thin},thin,line width=0.01, spy connection path={
        \draw [opacity=0.5] (tikzspyonnode.south west) -- (tikzspyinnode.south west);
        \draw [opacity=0.5] (tikzspyonnode.south east) -- (tikzspyinnode.south east);
        \draw [opacity=0.5] (tikzspyonnode.north west) -- (intersection of  tikzspyinnode.north west--tikzspyonnode.north west and tikzspyinnode.south east--tikzspyinnode.south west);
        \draw [opacity=0.5] (tikzspyonnode.north east) -- (intersection of  tikzspyinnode.north east--tikzspyonnode.north east and tikzspyinnode.south east--tikzspyinnode.south west);
        ;}] on (spypoint) in node at (spyviewer); %

        \addplot[mark=none, color=black,line width=\lwidth, opacity= \opac] table [x = ebn0, y = bp]{./figures/FG_273-191_20Iter.txt};
        \addplot[dashed,color=KITgreen!\opac!white,line width=\lwidth, opacity =\opac] table [x=ebn0, y=ddbmp]{./figures/FG_273-191_20Iter.txt};
        \addplot[dotted, mark=o,mark size =\marksz,every mark/.append style={solid, fill=black!\opac!white}, color=KITblue!\opac!white,line width=\lwidth, opacity =\opac] table [x = ebn0, y = srnms]{./figures/FG_273-191_20Iter.txt};
        \addplot[dotted, mark=x,mark size =\marksz,every mark/.append style={solid, fill=black!\opac!white}, color=KITblue!\opac!white,line width=\lwidth, opacity =\opac] table [x = ebn0, y = nms]{./figures/FG_273-191_20Iter.txt};
        \addplot[dotted,color=black!\opac!white,line width=\lwidth, opacity =\opac] table [x = ebn0, y = ms]{./figures/FG_273-191_20Iter.txt};
        
        \addplot[dashed,mark=none, color=KITorange, line width=\lwidth] table [x=ebn0, y=snn1]{./figures/FG_273-191_20Iter.txt};
        \addplot[dotted,color=KITorange, line width=\lwidth] table [x=ebn0, y=snn2_nflc]{./figures/FG_273-191_20Iter.txt};
        \addplot[solid,color=KITorange, line width=\lwidth] table [x=ebn0, y=snn2]{./figures/FG_273-191_20Iter.txt};
        
        \addlegendimage{} \addlegendentry{SPA}
        \addlegendimage{} \addlegendentry{DD-BMP}
        \addlegendimage{} \addlegendentry{SR-NMS}
        \addlegendimage{} \addlegendentry{NMS}
        \addlegendimage{} \addlegendentry{MS}
        
        \addlegendimage{} \addlegendentry{\Systemname\Discriminator}
        \addlegendimage{} \addlegendentry{\Systemname-$L_\mathrm{c}$}
        \addlegendimage{} \addlegendentry{\Systemname}

    \begin{pgfonlayer}{up}
        
    \end{pgfonlayer}
\end{axis}
\end{tikzpicture}} \\[-0.5em]
        \hspace*{1.2cm} \parbox{.75\textwidth}{\footnotesize (a) BER curve for the ($273$,$191$) FG LDPC code}. \\
    \end{minipage}
    \begin{minipage}{1\textwidth}
        \resizebox{.45\textwidth}{!}{\pgfplotsset{
layers/my layer set/.define layer set={
background,
main,
up
}{
 },
    set layers=my layer set,
}

\begin{tikzpicture}[spy using outlines={rectangle, magnification=2.7, size=1cm, connect spies}]
    \def\lwidth{1}
    \def\opac{70}
    \def\marksz{1.5pt}

    \def\spywidth{2.0cm}
    \def\spyheigth{1.5cm}

    \begin{axis}[
        ymode = log,
        xscale=1,
        xlabel style = {align=center},
        xlabel=$E_\mathrm{b}/N_0$ (dB),
        x label style={at={(axis description cs:0.5,0.02)},anchor=north},
        ylabel=BER,
        y label style={at={(axis description cs:0.00,0.4)},anchor=west},
        xtick = {0,1,2,3,4,5},
        xticklabels={$0$,$1$,$2$,$3$,$4$,$5$},
        grid=major,
        legend cell align={left},
        legend style={
            at={(0.02,0.01)},
            anchor=south west,
            fill opacity = 0.8,
            draw opacity = 1, 
            text opacity = 1,
        },
        xmin=2,xmax=4.2,
        ymin=7.6e-8,ymax=5e-1,
        axis line style=thick,
        tick label style={/pgf/number format/fixed},
        legend image post style={scale=2},
        ]

        \addplot[mark=none, color=black,line width=\lwidth,opacity= \opac] table [x = ebn0, y = bp]{./figures/FG_1023-781_20Iter.txt};
        \addplot[dashed, color=KITgreen!\opac!white,line width=\lwidth, opacity =\opac] table [x=ebn0, y=ddbmp]{./figures/FG_1023-781_20Iter.txt};
        \addplot[dotted,mark=o,every mark/.append style={solid, fill=black!\opac!white}, color=KITblue!\opac!white,line width=\lwidth, opacity =\opac] table [x = ebn0, y = srnms]{./figures/FG_1023-781_20Iter.txt};
        \addplot[dotted,mark=x,every mark/.append style={solid, fill=black!\opac!white}, color=KITblue!\opac!white,line width=\lwidth, opacity =\opac] table [x = ebn0, y = nms]{./figures/FG_1023-781_20Iter.txt};
        \addplot[dotted, color=black!\opac!white,line width=\lwidth, opacity =\opac] table [x = ebn0, y = ms]{./figures/FG_1023-781_20Iter.txt};
        
        \addplot[dashed, color=KITorange, line width=\lwidth] table [x=ebn0, y=snn_1]{./figures/FG_1023-781_20Iter.txt};
        \addplot[color=KITorange, line width=\lwidth] table [x=ebn0, y=snn_2]{./figures/FG_1023-781_20Iter.txt};
        
        \addlegendimage{} \addlegendentry{SPA}
        \addlegendimage{} \addlegendentry{DD-BMP}
        \addlegendimage{} \addlegendentry{SR-NMS}
        \addlegendimage{} \addlegendentry{NMS}
        \addlegendimage{} \addlegendentry{MS}
        
        \addlegendimage{} \addlegendentry{\Systemname\Discriminator}
        \addlegendimage{} \addlegendentry{\Systemname}

    \begin{pgfonlayer}{up}
        
    \end{pgfonlayer}
\end{axis}
\end{tikzpicture}} \\[-0.5em]%
        \hspace*{1.2cm} \parbox{.75\textwidth}{\footnotesize (b) BER curve for the ($1023$,$781$) FG LDPC code.}
    \end{minipage}
    \caption{BER curves of the \Systemname~, \Systemname\Discriminator~ and reference decoders.}
    \label{fig:results}
    \vspace{-0.7cm}
\end{figure}
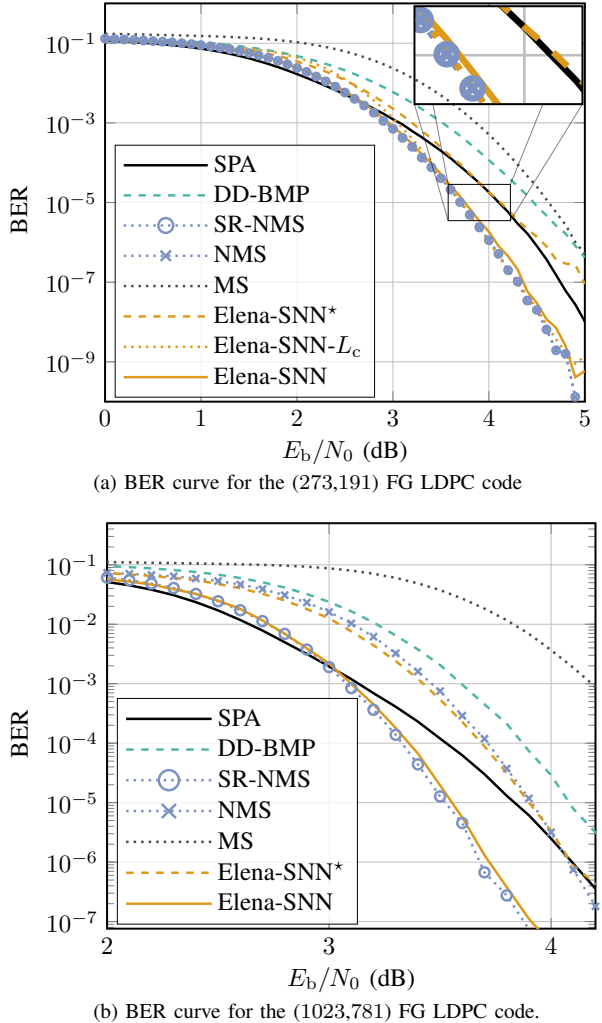

\vspace*{-0.1cm}
Fig.~\ref{fig:results} compares the different decoders for \Cs~and \Cb.
As expected, the SPA performs best in the low SNR regime for both codes.
For higher SNR, the SR-NMS decoder performs best.
Comparing the SR-NMS and the NMS decoders reveals that SR decreases the BER for \Cb.
The \Systemname\Discriminator~outperforms DD-BMP for both codes.
\Systemname~performs slightly worse than the SR-NMS decoder.
At a BER of $10^{-5}$, \Systemname~has a gap of $0.046$ dB to the SR-NMS decoder for \Cs. 
\Systemname~yields an SNR gain of $0.35\,\mathrm{dB}$ compared to SPA.
For \Cb, \Systemname~achieves a gap of $0.03\,\mathrm{dB}$ to SR-NMS and a gain of $0.28\,\mathrm{dB}$~over~SPA.
\RevI{
The \Systemname~decoder exhibits an error floor for $E_\mathrm{b}/N_0$ greater than $\SI{4.9}{\decibel}$, as Fig.~\ref{fig:results}(a) shows, resulting in a BER of approximately $10^{-9}$.
The error floor results from the low resolution of the approximation in \eqref{eq:elena}; future work will investigate an approximation with higher resolution and dynamic range.
For the \Cb~code, we simulated $10^{8}$ codewords at an $E_\mathrm{b}/N_0$ of $\SI{4.48}{\decibel}$ and did not observe an error floor.}
\RevII{Also, the constant $L_\mathrm{c}$ enables decoding without continuous SNR measurement, with improvement at low SNRs and only a slight penalty for high SNRs, as Fig.~\ref{fig:results}(a) shows.}

\vspace*{-0.4cm}
\subsection{Discussion}
\vspace*{-0.1cm}
We want to emphasize three significant points:
First, \Systemname~outperforms the SPA in high SNR regimes.
Compared to \Systemname, the simple \Systemname\Discriminator~decoder reduces the computational complexity at the cost of decoding performance.
It can only reach SPA performance; however, it solely requires the multiplication of signs and a single LI neuron.
Second, \Systemname~and \Systemname\Discriminator~were optimized for a fixed $E_\mathrm{s}/N_0$ and $L_\mathrm{c}$ value.
The results show that both approaches can generalize over a broad SNR regime. Thus, the set of applied parameters solely depends on the code, not the channel quality.
Third, in high SNR regimes, the variable-to-check-node messages $L_{i\rightarrow j}^{[\mathrm{v}]}$ tend to have large absolute values.
Hence, the LIF spiking neurons of the \Systemname~decoder are not sufficiently charged to generate an output spike.
\RevI{
Therefore, at $E_\mathrm{b}/N_0=3.5$ dB, a spike rate of $0.6\%$ is achieved, whereas, at $E_\mathrm{b}/N_0=0$ dB, the spike rate is increased to $9\%$ at $0$dB.
Hence, depending on the SNR, the number of emitted spikes varies for the code \Cs~between 557 spikes per codeword and 8354 spikes per codeword between $0$ dB and $3.5$ dB.}
\vspace*{-3mm}

\vspace*{-1.5mm}
\section{Conclusion}
\vspace*{-.5mm}
\label{sec:Conclusion}
In this work, we introduced \Systemname, a novel channel decoder for LDPC codes, which overcomes the complexity of the SPA by replacing its computationally complex calculations with an SNN.
We furthermore introduced a simplified version of the \Systemname~decoder, namely \Systemname\Discriminator, which significantly reduces the number of neurons.
The number of parameters that need to be optimized are three parameters for the \Systemname~decoder and one parameter for the \Systemname\Discriminator~decoder.
For the (273,191) and the (1023,781) FG LDPC codes, BPSK, and an AWGN channel, we have shown that both decoders outperform complexity-reduced versions of the SPA, e.g., MS and DD-BMP.
In high SNR regimes, the \Systemname~decoder outperforms SPA. 
Both decoders were optimized at a fixed $E_\mathrm{s}/N_0$ value and have been shown to generalize over the whole range of evaluated $E_\mathrm{s}/N_0$ values.
Hence, the decoder works without knowledge of the $E_\mathrm{s}/N_0$ of the system, which is an advantage compared to the SPA. 

Future work will test both proposed decoders on neuromorphic hardware and compare their complexity with the benchmark decoders.
Furthermore, we will evaluate the proposed decoders for different types of LDPC codes.
We are aware that the SNN input of the proposed decoders is real-valued. 
Thus, the neuromorphic hardware needs to support real-valued input, like, e.g., Intel Loihi \cite{Intel2021}.


\vspace*{-0.3cm}

\begin{thebibliography}{14}
\bibliographystyle{IEEEtran}

\bibitem{FSTNP2017}
T. Ferreira de Lima, B. Shastri, A. Tait, M. Nahmias, and P. Prucnal, ``Progress in neuromorphic photonics,'' \textit{Nanophotonics}, vol. 6, no. 3, pp. 577-599, 2017. 

\bibitem{RJP2019}
K. Roy, A. Jaiswal, and P. Panda, ``Towards spike-based machine intelligence with neuromorphic computing,'' \textit{Nature}, vol. 575, pp. 607–617, 2019

\bibitem{YK2020}
Y. S. Yang, and Y. Kim, ``Recent trend of neuromorphic computing hardware: Intel's neuromorphic system perspective,'' \textit{Proc. International SoC Design Conference (ISOCC)}, Yeosu, Korea (South), 2020, pp. 218-219.

\bibitem{Intel2021}
Intel, ``Taking neuromorphic computing to the next level with Loihi 2,'' 2021. [Online]. Available: \url{https://download.intel.com/newsroom/2021/new-technologies/neuromorphic-computing-loihi-2-brief.pdf}, (accessed on: 22.05.24).

\bibitem{PBXKSSWLMS22}  
  C. Pehle \textit{et al.}, ``The {BrainScaleS-2} accelerated neuromorphic system with hybrid plasticity,'' \textit{Front. Neurosci.}, vol.16, Feb. 2022. 

\RevI{\bibitem{MNHW24} M. Moursi, J. Ney, B. Hammoud, and N. Wehn, ``Efficient FPGA Implementation of an Optimized SNN-based DFE for Optical Communications'', \textit{IEEE Middle East Conference on Communications and Networking (MECOM 2024)}, November, 2024, Abu Dhabi, United Arab Emirates
}

\bibitem{KLF2001}
Y. Kou, S. Lin and M. P. C. Fossorier, ``Low-density parity-check codes based on finite geometries: a rediscovery and new results,'' in \textit{IEEE Trans. Inf. Theory}, vol. 47, no. 7, pp. 2711-2736, Nov. 2001.

\bibitem{AVAT2014}
F. Angarita, J. Valls, V. Almenar and V. Torres, ``Reduced-complexity min-sum algorithm for decoding LDPC codes with low error-floor,'' in \textit{IEEE Trans. Circuits Syst. I: Regul. Pap.}, vol. 61, no. 7, pp. 2150-2158, July 2014.

\bibitem{MBH2009}
N. Mobini, A. H. Banihashemi and S. Hemati, ``A differential binary message-passing LDPC decoder,'' in \textit{IEEE Trans. Commun.}, vol. 57, no. 9, pp. 2518-2523, Sep. 2009.

\bibitem{XTB2008}
H. Xiao, S. Tolouei and A. H. Banihashemi, ``Successive relaxation for decoding of LDPC codes,'' \textit{Proc. Biennial Symposium on Communications}, Kingston, ON, CA, 2008.

\bibitem{JB2012}
E. Janulewicz and A. H. Banihashemi, ``Performance analysis of iterative decoding algorithms with memory over memoryless channels,'' in \textit{IEEE Trans. Commun.}, vol. 60, no. 12, pp. 3556-3566, Dec. 2012.

\bibitem{CDEFX2005}
J. Chen, A. Dholakia, E. Eleftheriou, M. P. C. Fossorier and X.-Y. Hu, ``Reduced-complexity decoding of LDPC codes,'' in \textit{IEEE Trans. Commun.}, vol. 53, no. 8, pp. 1288-1299, Aug. 2005.

\bibitem{SMD09}
A. Steimer, W. Maass and R. Douglas, ``Belief propagation in networks of spiking neurons,'' \textit{Neural Comput.}, vol. 21, no. 9, pp. 2502-23, Sep. 2009.

\bibitem{github-code}
A. von Bank,  ``Enlarge-Likelihood-Each-Notable-Amplitude Spiking-Neural-Network (ELENA-SNN)
decoder,'' 2024. [Online]. Available: \url{https://github.com/kit-cel/ELENA}

\bibitem{NMZ2019}
E. O. Neftci, H. Mostafa and F. Zenke, ``Surrogate gradient learning in spiking neural networks: bringing the power of gradient-based optimization to spiking neural networks,'' in \textit{IEEE Signal Process. Mag.}, vol. 36, no. 6, pp. 51-63, Nov. 2019.

\bibitem{Norse}
C. Pehle and J. E. Pedersen, ``Norse - A deep learning library for spiking
neural networks,'' Jan. 2021, documentation: https://norse.ai/docs/.   	

\bibitem{RU2008}
T. Richardson and R. Urbanke, \textit{Modern Coding Theory}, Cambridge
University Press, 2008.

\end{thebibliography}
\end{document}